\documentstyle[twoside,fleqn,espcrc2,epsf,epsfig]{article}

\newcommand{\Dslash}{$D$\kern-0.6em \hbox{/}}

\newcommand{\la}{\raise.16ex\hbox{$\langle$}}
\newcommand{\ra}{\raise.16ex\hbox{$\rangle$}}
\newcommand{\be}{\begin{equation}}
\newcommand{\ee}{\end{equation}}
\newcommand{\bea}{\begin{eqnarray}}
\newcommand{\eea}{\end{eqnarray}}
\newcommand{\noi}{\noindent}
\newcommand{\pst}{\protect\textstyle\scriptscriptstyle}

\title{Improving the sign problem in QCD at finite density
\thanks{Talk presented by V. Laliena}}

\author{Ph. de Forcrand and 
V. Laliena\address{ETH, CH-8092 Z\"urich, Switzerland} }

\begin{document}

\begin{abstract}
If the fermion mass is large enough, the phase of the fermion determinant of
QCD at finite density is strongly correlated with the imaginary part of the 
Polyakov loop. This fact can be exploited to reduce the fluctuations of the 
phase significantly , making numerical simulations feasible in regions
of parameters where the naive brute force method does not work.
\end{abstract}
\maketitle

\vspace*{-0.9cm}

\section{The Polyakov loop and the phase of the determinant}

It is well known that the euclidean path integral representation of the
QCD partition function at finite chemical potential suffers from the so-called
sign problem: the fermion determinant is complex and the theory
cannot be simulated with the usual Monte Carlo method. One could still try 
to get results with the simple brute force method, that is, simulate a
positive measure which contains the pure gauge action and the modulus of the
determinant, treating the phase as an observable. Then, the expectation value
of any observable $\cal{O}$ is given by the ratio
$\langle {\cal O}\rangle = 
\langle {\cal O} \exp i\theta\rangle_m / 
\langle \cos\theta\rangle_m$,
where $\langle \cdot \rangle_m$ denotes the expectation value using 
the modulus of the determinant as a probability measure, and $\theta$ is the 
phase of the determinant (PD). Unfortunately, the expectation value
of $\cos\theta$ is a positive quantity 
exponentially small with the volume. Since the relative error 
$\langle {\cal O}\rangle$ is given by
\be
\frac{\delta \langle {\cal O}\rangle}{\langle {\cal O}\rangle} = 
\frac{\delta \langle {\cal O}\exp i\theta\rangle_m}
{\langle {\cal O}\exp i\theta\rangle_m} + 
\frac{\delta \langle \cos\theta\rangle_m}{\langle \cos\theta\rangle_m}\, ,
\label{relerr}
\ee
\noi
$\langle\cos\theta\rangle_m$ must be measured very accurately to achieve 
a given accuracy for $\langle {\cal O}\rangle$.  This requires statistic 
growing exponentially with the volume. Obviously, this is, from the
numerical point of view, an almost hopeless task.

Due to the up to now unsurmountable difficulties with QCD with light quarks
at finite density, increasing attention is being paid to the limit of
infinitely heavy quarks \cite{bht}-\cite{nos}. The sign problem still remains
in this limit, but numerical computations are easier \cite{bht}. For large
quark mass $m$, the logarithm of the fermion determinant can be expanded in 
powers of $1/(2m)$ (we focus our discussion on staggered fermions).
The first term sensitive to the chemical potential $\mu$ is
\be
\left(\frac{e^\mu}{2m}\right)^{N_T}\,\sum_{\vec{x}} {\rm Tr} L_{\vec{x}}
\:+\: \left(\frac{e^{-\mu}}{2m}\right)^{N_T}\,
\sum_{\vec{x}}{\rm Tr} L_{\vec{x}}^\dagger\, ,
\ee
\noi
where $N_T$ is the number of lattice points in
the temporal direction, $\vec{x}$ labels the sites in a given temporal
slice, and $L_{\vec{x}}$ is the ordered product of all temporal links attached
to $\vec{x}$. Hence, to this order, the PD is 
proportional to the imaginary part of the Polyakov loop (IPPL): 
$\theta= c V_s L_i$, with $L_i=1/V_s \sum_{\vec{x}} {\rm Im Tr} L_{\vec{x}}$ 
and $c=2\sinh(\mu N_T)/(2m)^{N_T}$. Of course, higher order corrections will 
destroy this linear relationship.

The static limit ($m,\mu\rightarrow\infty$ with 
$c=[\exp(\mu)/2m]^{N_T}$ fixed) has been
 studied in \cite{bht}. Analyzing the data of \cite{bht}, we found
a strong correlation between the PD and the IPPL, 
as can be seen in Figs. 1 and 2 \cite{nos}. The strong correlation displayed 
in Fig. 1 is not surprising, since the data correspond to $c\approx 0.05$, 
which is small enough for the linear relation previously discussed to become 
essentially exact. Fig. 2 is more interesting, since $c\approx 1.66$ is not 
small. There, we 
can see that the linear correlation still holds, though the width of the band 
is much larger than in Fig. 1. Very recently, a paper confirming these 
findings from a continuum analysis has appeared \cite{langfeld}.

\newpage

\begin{figure}[t]
\begin{center}
\epsfig{file=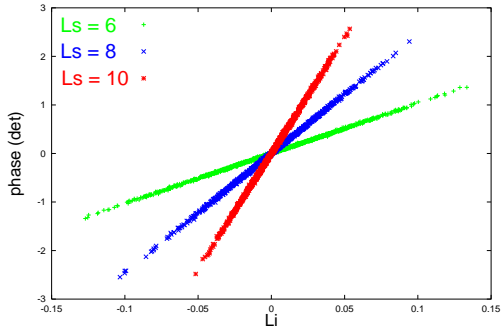,height=4.5cm,width=6.5cm,angle=-90}
\end{center}
\vspace*{-1.5 cm}
\caption{Correlation between The PD $\theta$ and The IPPL $L_i$ in static QCD, 
with $C=1/20$ and $\beta=5.0$ on a lattice $L_s^3\times 2$. The data are from
Ref. 1.}
\end{figure} 

\vspace*{-1.5cm}

\section{Fixing the imaginary part of the Polyakov loop}

Given the correlation between the PD and the IPPL, it is plausible that, at 
least for heavy enough quarks, the fluctuations of the PD
would be suppressed to a large extent provided we constrain our path integral
to configurations with real Polyakov loop. To see whether this is possible 
we write the partition function as
${\cal Z} = \int dp_i\,\exp\left(-V_s {\cal F}(p_i)\right)$, where $V_s$ is
the spatial lattice volume and 
\be
e^{-V_s{\cal F}}=\int[dU]e^{-S_g(U)}\det\Delta(U)\,\delta(p_i-L_i)\, .
\ee
\noi
For $V_s\rightarrow\infty$ the integral in $p_i$ is saturated by the saddle
point $p_i^{\pst sp}$, which is in general complex, since ${\cal F}$ is. 

It is easy to show that the saddle point is indeed purely imaginary. The 
expectation value of the IPPL coincides with $p_i^{\pst sp}$, and therefore
\be
p_i^{\pst sp} = 
\frac{\langle L_i\cos\theta\rangle_m}{\langle\cos\theta\rangle_m} +
i\,\frac{\langle L_i\sin\theta\rangle_m}{\langle\cos\theta\rangle_m} \, .
\label{expect}
\ee
\noi
For each gauge configuration we also have its complex conjugate, for which
$L_i$ changes sign. From the loop expansion of the logarithm of the fermion
determinant, we see that the modulus and the phase depend only on the real 
and imaginary part of the loops respectively. Hence, the modulus and the
pure gauge action do not change, while $\theta$ changes sign. The first
term in the r.h.s. of (\ref{expect}) vanishes, so that $p_i^{\pst sp}$ is
purely imaginary.

The existence of a saddle point for the partition function implies the 
equivalence between canonical and microcanonical ensembles. One can constrain
an observable to its saddle point value. This only makes sense when
the saddle point is real.
Therefore, we can only constrain the IPPL to zero in those cases where
its expectation value is zero. We expect this to happen at zero temperature 
\cite{nos}. At finite temperature, the expectation value of the Polyakov
loop gives the free energy of a heavy quark, and its complex conjugate
that of a heavy antiquark. Since
these two free energies should be different, the expectation value of the
IPPL cannot be zero.
However, if the temperature is small, the expectation value of IPPL should
be exponentially small with $1/T$. The static limit at strong coupling, which
can be solved analytically \cite{zara}, confirms this. In this case, 
$p_i^{\pst sp} = (c^2 - c)/(c^3+1)$, with $c=(e^\mu/2m)^{N_T}$, in
agreement with our expectation. Therefore, we can constrain the IPPL to zero 
in simulations of the low temperature phase of QCD at finite density.

\begin{figure}[t]
\begin{center}
\epsfig{file=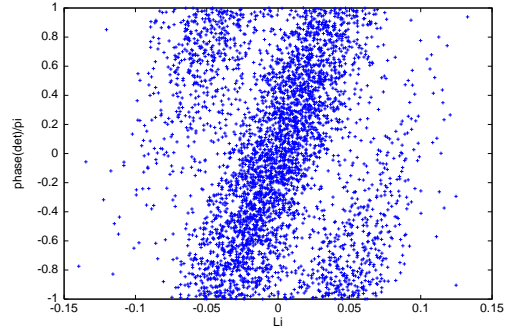,height=4.5cm,width=6.5cm}
\end{center}
\vspace*{-1.5 cm}
\caption{Same as Fig. 1 with
$C\approx 1.66$ on a $6^3\times 4$ lattice.} 
\end{figure} 

\section{Constrained Monte Carlo}

To test our ideas we developed a HMC algorithm in which
the molecular dynamics is forced to evolve on the surface of zero IPPL.
This is easily achieved by introducing a Lagrange multiplier which must
be computed at each step of the molecular dynamics by solving a non-linear
equation, which is the condition that the constraint must be obeyed at each
step. It can be shown that the dynamics is reversible and that 
detailed balance is satisfied. The additional cost in CPU time 
caused by the constraint is small: 10\% more than the unconstrained case
for quenched simulations, and completely negligible with dynamical fermions.

As a preliminary test, we made several quenched runs to check the gain of
a constrained simulation in comparison with an unconstrained one.
We worked with a $4^3\times 6$ lattice at $\beta=1.0$ to ensure that we were
in the low temperature phase. We got a set of 300 equilibrium quenched 
configurations and we diagonalized its associated massless
staggered fermion matrix, for $\mu=0.5,1.0,1.5$ and $2.0$. Having the 
eigenvalues of the massless matrix, it is very cheap to get the determinant 
for any value of the mass. 

Let us describe our results. If the mass is large, there is no sign problem
in the small and large $\mu$ regions. Simulations are easy but not interesting
there. When $\mu$ is of the order $m$ the sign problem becomes severe: this
is the region where the onset transition, separating the zero from the finite
density phases, takes place. It is very interesting
to determine it accurately. Fig. 3a displays $\langle\cos\theta\rangle$ as
a function of $m$, for $\mu=1.5$. The logarithmic scale
allows us to see the relative error entering Eq. (\ref{relerr}).
In some cases, the relative error in the unconstrained simulation is one
order of magnitude larger than in the constrained one.
Notice the severe sign
problem signaling a transition for $m\sim 1.75$. The transition
region is very broad in the unconstrained simulation, and in fact covers most
of the finite density region, between saturation (low $m$) and zero density 
(high $m$). 
Actually, almost nothing can be inferred about the finite density phase in the
unconstrained case. In the constrained simulation,
the sign problem occurs in a much narrower window, so that the onset 
transition as well as the finite density phase could be analysed.
From Fig. 3a, we see that $\mu=1.5$ is nearly critical for 
$m\in [1.75,1.82]$. As expected, the constraint in the IPPL
improves the sign problem and makes numerical simulations feasible.
Fig. 3b displays the distribution of $\theta/\pi$ for $\mu=1.5$ and
$m=1.5$. The difference between the constrained and unconstrained cases is
manifest.
An extended version of this work has recently appeared \cite{nos}.

We thank the authors of Ref. \cite{bht}, especially Doug Toussaint,
for making their data available to us. Ph. de F. thanks Mike Creutz for 
helpful discussions. V.L. acknowledges useful discussions with R. Aloisio,
V. Azcoiti and A. Galante.

%\vspace*{-1. cm}
\begin{figure}[t]
\begin{center}
\epsfig{file=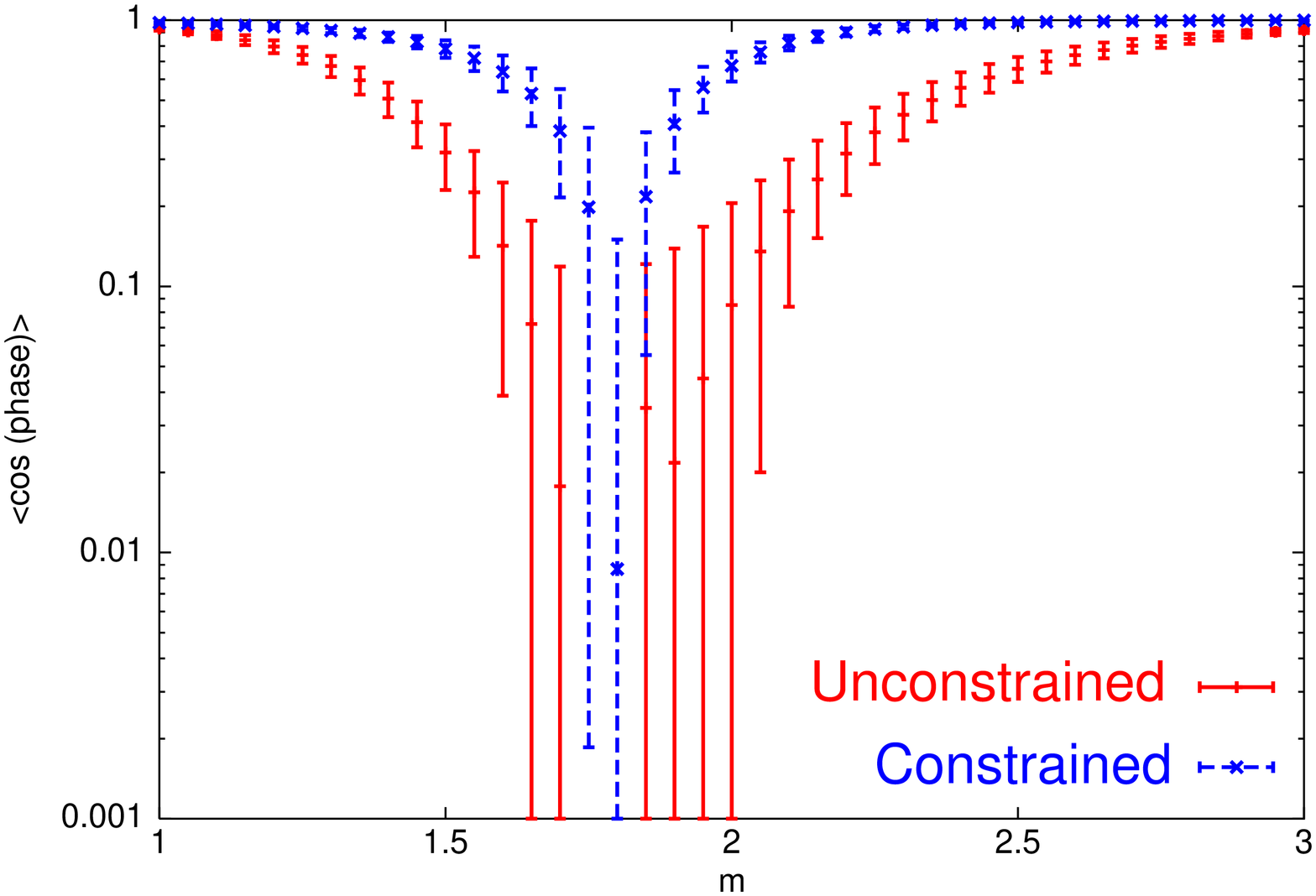,height=4.5cm,width=7.0cm,angle=-90}
\epsfig{file=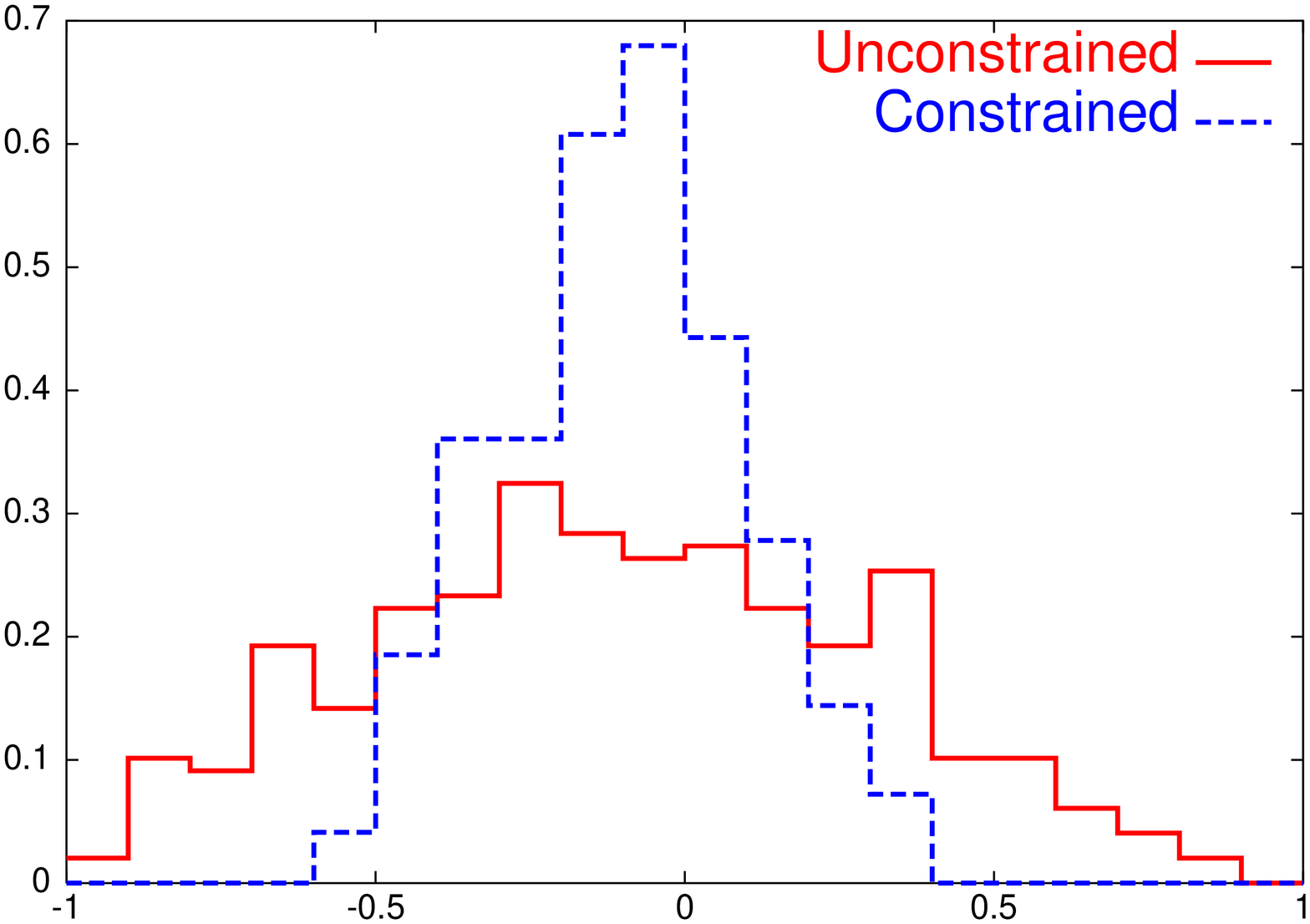,height=4.5cm,width=7.0cm,angle=-90}
\end{center}
\vspace{-1.5 cm}
\caption{a) Average of $\cos\theta$ vs. fermion mass for $\mu=1.5$;
b) Histogram of $\theta/\pi$ for $\mu=1.5$ and $m=1.5$ .}
\end{figure}

\end{document}